\documentclass[aps,prd,onecolumn,showpacs,showkeys,amsmath,amssymb]{revtex4}
\usepackage{amsfonts}
\usepackage{amsmath}
\usepackage{graphicx}
\usepackage{subfig}
\usepackage{dcolumn}
\usepackage{natbib}
\usepackage{bm}
\usepackage{booktabs}
\usepackage{slashed}

\usepackage{ulem}
\usepackage{color}

\makeatletter
\newcommand{\figcaption}{\def\@captype{figure}\caption}
\newcommand{\tabcaption}{\def\@captype{table}\caption}

\newcommand{\Rmnum}[1]{\expandafter\@slowromancap\romannumeral #1@}
\def\hlinewd#1{%
  \noalign{\ifnum0=`}\fi\hrule \@height #1 \futurelet
   \reserved@a\@xhline}
\makeatother
\def\dab{\int^{\alpha_{max}}_{\alpha_{min}}d\alpha\int^{\beta_{max}}_{\beta_{min}}d\beta}
\def\qq{\langle\bar qq\rangle}
\def\ss{\langle \bar ss\rangle}
\def\GGa{\langle GG\rangle}
\def\GGb{\langle g_s^2GG\rangle}
\def\qGqa{\langle\bar qg_s\sigma\cdot Gq\rangle}
\def\qGqb{\langle\bar qGq\rangle}
\def\sGsa{\langle\bar sg_s\sigma\cdot Gs\rangle}

\def\f(s){\left[(\alpha+\beta)m_c^2-\alpha\beta s\right]}
\def\non{\\ \nonumber}

\begin{document}
%
%
\title{Mass spectra for $qc\bar q\bar c$, $sc\bar s\bar c$, $qb\bar q\bar b$, $sb\bar s\bar b$
tetraquark states with $J^{PC}=0^{++}$ and $2^{++}$}

 \author{Wei Chen$^{1,2}$}
 \author{Hua-Xing Chen$^3$}
 \email{hxchen@buaa.edu.cn}
 \author{Xiang Liu$^{4,5}$}
 \email{xiangliu@lzu.edu.cn}
 \author{T. G. Steele$^1$}
 \email{tom.steele@usask.ca}
 \author{Shi-Lin Zhu$^{6,7,8}$}
 \email{zhusl@pku.edu.cn}

\affiliation{
$^1$School of Physics, Sun Yat-Sen University, Guangzhou 510275, China
\\
$^2$Department of Physics and Engineering Physics, University of Saskatchewan, Saskatoon, Saskatchewan, S7N 5E2, Canada
\\
$^3$School of Physics and Beijing Key Laboratory of Advanced Nuclear Materials and Physics, Beihang University, Beijing 100191, China
\\
$^4$School of Physical Science and Technology, Lanzhou University, Lanzhou 730000, China
\\
$^5$Research Center for Hadron and CSR Physics, Lanzhou University and Institute of Modern Physics of CAS, Lanzhou 730000, China
\\
$^6$School of Physics and State Key Laboratory of Nuclear Physics and Technology, Peking University, Beijing 100871, China
\\
$^7$Collaborative Innovation Center of Quantum Matter, Beijing 100871, China
\\
$^8$Center of High Energy Physics, Peking University, Beijing 100871, China
}

\begin{abstract}
We have studied the mass spectra of the hidden-charm/bottom $qc\bar
q\bar c$, $sc\bar s\bar c$ and $qb\bar q\bar b$, $sb\bar s\bar b$
tetraquark states with $J^{PC}=0^{++}$ and $2^{++}$ in the framework
of QCD sum rules. We construct ten scalar and four tensor
interpolating currents in a systematic way and calculate the mass
spectra for these tetraquark states. The $X^\ast(3860)$ may be either an
isoscalar tetraquark state or $\chi_{c0}(2P)$. If the $X^\ast(3860)$
is a tetraquark candidate, our results prefer the $0^{++}$ option
over the $2^{++}$ one. The $X(4160)$ may be classified as either
the scalar or tensor $qc\bar q\bar c$ tetraquark state while the
$X(3915)$ favors a $0^{++}$ $qc\bar q\bar c$ or $sc\bar s\bar c$
tetraquark assignment over the tensor one. The $X(4350)$ can
not be interpreted as a $sc\bar s\bar c$ tetraquark with either
$J^{PC}=0^{++}$ or $2^{++}$.
\end{abstract}


 \maketitle
%
\section{Introduction}\label{Sec:Intro}

In $B$ factories, the two photon fusion process $\gamma\gamma\to X$
is used to produce $C$-even charmonium states. To date, the Belle
Collaboration have reported three charmnomium-like states in this
process. They are the $Z(3930)$ state in the $\gamma\gamma\to D\bar
D$~\cite{2006-Uehara-p82003-82003}, the $X(3915)$ state in
$\gamma\gamma\to\omega J/\psi$
process~\cite{2010-Uehara-p92001-92001} and the $X(4350)$ state in
the $\gamma\gamma\to\phi J/\psi$
process~\cite{2010-Shen-p112004-112004}. Since these three states
were produced in the $\gamma\gamma$ fusion process, their possible
quantum numbers can be either $J^{PC}=0^{++}$ or $2^{++}$.

In 2008, Belle analyzed the double charmonium production $e^+e^-\to
J/\psi D^{\ast+}D^{\ast-}$ process and found a new charmonium-like
structure $X(4160)$ with a significance of
$5.1\sigma$~\cite{2008-Pakhlov-p202001-202001}. At present, the
$D^{\ast+}D^{\ast-}$ is the only observed decay mode for the
$X(4160)$ state. If $e^+e^-\to J/\psi X(4160)$ is dominant by
$e^+e^-\to\gamma^\ast\to J/\psi X(4160)$, the $C$-parity of
$X(4160)$ should be positive. Very recently, Belle performed a full
amplitude analysis of the double charmonium production process
$e^+e^-\to J/\psi D\bar D$ and observed a new charmonium-like
structure $X^\ast(3860)$ with a significance of $6.5
\sigma$~\cite{2017-Chilikin-p112003-112003}. Using Monte Carlo simulation, Belle compared
the $J^{PC}=0^{++}$ and $2^{++}$ hypotheses for the $X^\ast(3860)$
and found that the $J^{PC}=0^{++}$ hypothesis is favored, although
the $2^{++}$ hypothesis is not excluded~\cite{2017-Chilikin-p112003-112003}.

The masses and decay widths for the $X(4160)$, $Z(3930)$, $X(3915)$,
$X(4350)$ and $X^\ast(3860)$ are shown in Table~\ref{states}. Their
possible quantum numbers are also listed in the second column.
According to the GI (Godfrey-Isgur) model
calculations~\cite{1985-Godfrey-p189-231,2005-Barnes-p54026-54026},
the $Z(3930)$ has been assigned as the $2^3P_2$ radially excited
charmonium $\chi^\prime_{c2}(2P)$ with $J^{PC}=2^{++}$ while the
$X(3915)$ as $\chi_{c0}(2P)$ charmonium state with $J^{PC}=0^{++}$
in PDG~\cite{2016-Patrignani-p100001-100001}. Such an assignment was
also supported by analyzing the mass spectrum of the P-wave
charmonium family and open-charm strong decay of the
$X(3915)$~\cite{2010-Liu-p122001-122001,2013-Jiang-p1350145-1350145}.
However, the $\chi_{c0}(2P)$ interpretation for $X(3915)$ was
challenged by the absence of the $D\bar D$ decay mode and small mass
splitting between $X(3915)$ and $Z(3930)$ compared with that between
$\chi_{c0}(2P)$ and $\chi_{c2}(2P)$~\cite{2011-Guo-p34013-34013}. In
Ref.~\cite{2017-Chilikin-p112003-112003}, Belle thus suggested the
$X^\ast(3860)$ as a better candidate for the $\chi_{c0}(2P)$
charmonium state than $X(3915)$ since its mass and decay mode are
well matched with the expectations for $\chi_{c0}(2P)$. This
suggestion was studied in a Friedrichs-model-like scheme in
Ref.~\cite{2017-Zhou-p-}. Additionally, the tetraquark interpretation was
also proposed to study the nature of $X(3915)$ and $X^\ast(3860)$.
In Ref.~\cite{2016-Lebed-p94024-94024}, the $X(3915)$ was considered
as the lightest $0^{++}$ $cs\bar c\bar s$ tetraquark state in the
diquark model. Such an interpretation was supported by the QCD sum
rule calculation~\cite{2017-Wang-p78-78}.
See also QCD sum rule studies in Refs.~\cite{Albuquerque:2009ak,Zhang:2009st,Wang:2009ue,Wang:2009ry,Chen:2016oma}.
The $X^\ast(3860)$ was
explained to be the scalar $cs\bar c\bar s$ state in
Refs.~\cite{2017-Wang-p-,2017-Yu-p-}.

\begin{table}
\renewcommand{\arraystretch}{1.3}
\begin{tabular*}{13.2cm}{ccccc}
\hlinewd{.8pt}
State  &  $J^{PC}$ & Process  &  Mass (MeV)  &  Width (MeV)  \\
\hline
$Z(3930)$~\cite{2006-Uehara-p82003-82003} &$2^{++}$& $\gamma\gamma\to D\bar D$ & $3929\pm5\pm2$ & $29\pm10\pm2$ \\
$X(3915)$~\cite{2010-Uehara-p92001-92001} &$0^{++}$ or $2^{++}$& $\gamma\gamma\to\omega J/\psi$ & $3915\pm3\pm2 \mbox{MeV}$ & $17\pm10\pm3$ \\
$X(4350)$~\cite{2010-Shen-p112004-112004} &$0^{++}$ or $2^{++}$& $\gamma\gamma\to\phi J/\psi$ & $4350.6^{+4.6}_{-5.1}\pm0.7 \mbox{MeV}$ & $13^{+18}_{-9}\pm4$ \\
$X(4160)$~\cite{2008-Pakhlov-p202001-202001} &$?^{?+}$& $e^+e^-\to J/\psi D^{\ast+}D^{\ast-}$ & $4156^{+29}_{-25}$ & $37^{+27}_{-17}$ \\
$X^\ast(3860)$~\cite{2017-Chilikin-p112003-112003} &$0^{++}$(prefered) or $2^{++}$& $e^+e^-\to J/\psi D\bar D$ & $3862^{+26+40}_{-32-13}$ & $201^{+154+88}_{-67-82}$ \\
\hlinewd{.8pt}
\end{tabular*}
\caption{Experimental parameters for $X(4160)$, $Z(3930)$,
$X(3915)$, $X(4350)$ and $X^\ast(3860)$.
\label{states}}
\end{table}

Since the $X(4160)$ was only observed in the $D^*\bar D^*$ final
states~\cite{2008-Pakhlov-p202001-202001}, its $J^P$ quantum numbers
has not been determined up to now. In
Ref.~\cite{2008-Chao-p348-353}, Chao ruled out the interpretations
of the $X(4160)$ as the $\psi(4160)$ or D-wave charmonium state $2
^1D_2$ with $J^{PC}=2^{-+}$ based on NRQCD calculations and proposed
the $X(4160)$ as a candidate of the $\eta_c(4S)$. However, the
$\eta_c(4S)$ assignment for the $X(4160)$ was in conflict with the
mass and decay width predictions for $\eta_c(4S)$
state~\cite{2005-Barnes-p54026-54026,2009-Li-p94004-94004,2014-He-p3208-3208}.
The $X(4160)$ was also explained as an isoscalar $D_s^*\bar D_s^*$
molecular state with $J^{PC}=2^{++}$ within the framework of the
hidden gauge formalism in Ref.~\cite{2009-Molina-p114013-114013}.
See also discussions in Refs.~\cite{Liang:2015twa,Guo:2015daa,Dai:2015bcc}.

In the recent reviews~\cite{2016-Chen-p1-121,2016-Esposito-p1-97,2017-Lebed-p143-194,2017-Guo-p-,Chen:2016spr}, one can
consult the latest progress on the $X(4160)$, $X(3915)$, $X(4350)$
and $X^\ast(3860)$ states. The tetraquark configuration is an
interesting explanation of their underlying structure. As shown in
Table~\ref{states}, the quantum numbers for the $X(4160)$,
$X(3915)$, $X(4350)$ and $X^\ast(3860)$ states can be
$J^{PC}=0^{++}$ or $2^{++}$. In this paper, we shall study the mass
spectra for the $qc\bar q\bar c$, $sc\bar s\bar c$, $qb\bar q\bar b$
and $sb\bar s\bar b$ tetraquark states with $J^{PC}=0^{++}$ and
$2^{++}$ in the method of QCD sum rules.

This paper is organized as follows. In Sect.~\Rmnum{2}, we systematically construct
the $qc\bar q\bar c$ tetraquark interpolating currents with
$J^{PC}=0^{++}$ and $2^{++}$ and introduce the
QCD sum rule formalism. Then we derive the spectral densities with
the two-point correlation functions. In Sect.~\Rmnum{3}, we perform
the QCD sum rule analyses and extract the mass spectra of the
$qc\bar q\bar c$, $sc\bar s\bar c$, $qb\bar q\bar b$ and $sb\bar
s\bar b$ tetraquark states. The last section is a brief discussion
and summary.

 \section{Formalism of QCD sum rules}\label{Sec:QSR}

To explore the charmonium-like tetraquark systems, we construct the
$qc\bar q\bar c$ diquark-antidiquark operators using the
following diquark fields $q^T_a Cc_b$, $q^T_a C\gamma_5c_b$, $q^T_a
C\gamma_\mu c_b$, $q^T_a C\gamma_\mu\gamma_5c_b$, $q^T_a
C\sigma_{\mu\nu}c_b$ with various Lorentz
structures~\cite{Chen:2007xr,2013-Du-p14003-14003,2014-Chen-p54037-54037,2011-Chen-p34010-34010,2010-Chen-p105018-105018}.
Using SU(3) color symmetry, we obtain the scalar interpolating
currents with quantum numbers $J^{PC}=0^{++}$
\begin{equation}
\begin{split}
J_1&=q^T_aC\gamma_5c_b(\bar{q}_a\gamma_5C\bar{c}^T_b+\bar{q}_b\gamma_5C\bar{c}^T_a)\, , \\
J_2&=q^T_aC\gamma_\mu c_b(\bar{q}_a\gamma^\mu C\bar{c}^T_b+\bar{q}_b\gamma^\mu C\bar{c}^T_a)\, ,\\
J_3&=q^T_aC\gamma_5c_b(\bar{q}_a\gamma_5C\bar{c}^T_b-\bar{q}_b\gamma_5C\bar{c}^T_a)\, ,\\
J_4&=q^T_aC\gamma_\mu c_b(\bar{q}_a\gamma^\mu C\bar{c}^T_b-\bar{q}_b\gamma^\mu C\bar{c}^T_a)\, , \\
J_5&=q^T_aCc_b(\bar{q}_aC\bar{c}^T_b+\bar{q}_bC\bar{c}^T_a)\, , \\
J_6&=q^T_aC\gamma_\mu\gamma_5 c_b(\bar{q}_a\gamma^\mu\gamma_5 C\bar{c}^T_b+\bar{q}_b\gamma^\mu\gamma_5 C\bar{c}^T_a)\, ,\\
J_7&=q^T_aC\sigma_{\mu\nu}c_b(\bar{q}_a\sigma^{\mu\nu}C\bar{c}^T_b+\bar{q}_b\sigma^{\mu\nu}C\bar{c}^T_a)\, ,\\
J_8&=q^T_aCc_b(\bar{q}_aC\bar{c}^T_b-\bar{q}_bC\bar{c}^T_a)\, , \\
J_9&=q^T_aC\gamma_\mu\gamma_5 c_b(\bar{q}_a\gamma^\mu\gamma_5 C\bar{c}^T_b-\bar{q}_b\gamma^\mu\gamma_5 C\bar{c}^T_a)\, ,\\
J_{10}&=q^T_aC\sigma_{\mu\nu}c_b(\bar{q}_a\sigma^{\mu\nu}C\bar{c}^T_b-\bar{q}_b\sigma^{\mu\nu}C\bar{c}^T_a)\,
,  \label{scalarcurrents}
\end{split}
\end{equation}
and the tensor interpolating currents with quantum numbers
$J^{PC}=2^{++}$
\begin{equation}
\begin{split}
J_{11\mu\nu}&=q^T_aC\gamma_\mu c_b(\bar{q}_a\gamma_\nu
C\bar{c}_b^T-\bar{q}_b\gamma_\nu C\bar{c}_a^T)
+q^T_aC\gamma_\nu c_b(\bar{q}_a\gamma_\mu C\bar{c}_b^T-\bar{q}_b\gamma_\mu C\bar{c}_a^T)\, , \\
J_{12\mu\nu}&=q^T_aC\gamma_\mu\gamma_5
c_b(\bar{q}_a\gamma_\nu\gamma_5
C\bar{c}_b^T-\bar{q}_b\gamma_\nu\gamma_5 C\bar{c}_a^T)
+q^T_aC\gamma_\nu\gamma_5 c_b(\bar{q}_a\gamma_\mu\gamma_5 C\bar{c}_b^T-\bar{q}_b\gamma_\mu\gamma_5 C\bar{c}_a^T)\, , \\
J_{13\mu\nu}&=q^T_aC\gamma_\mu c_b(\bar{q}_a\gamma_\nu
C\bar{c}_b^T+\bar{q}_b\gamma_\nu C\bar{c}_a^T)
+q^T_aC\gamma_\nu c_b(\bar{q}_a\gamma_\mu C\bar{c}_b^T+\bar{q}_b\gamma_\mu C\bar{c}_a^T)\, , \\
J_{14\mu\nu}&=q^T_aC\gamma_\mu\gamma_5
c_b(\bar{q}_a\gamma_\nu\gamma_5
C\bar{c}_b^T+\bar{q}_b\gamma_\nu\gamma_5 C\bar{c}_a^T)
+q^T_aC\gamma_\nu\gamma_5 c_b(\bar{q}_a\gamma_\mu\gamma_5
C\bar{c}_b^T+\bar{q}_b\gamma_\mu\gamma_5 C\bar{c}_a^T)\, ,
\label{tensorcurrents}
\end{split}
\end{equation}
in which the currents $J_{1}(x)$, $J_{2}(x)$, $J_{5}(x)$,
$J_{6}(x)$, $J_{7}(x)$, $J_{13\mu\nu}(x)$, $J_{13\mu\nu}(x)$ belong
to the $[\mathbf{6_c}]_{qc} \otimes [\mathbf{\bar 6_c}]_{\bar{q}\bar
c}$ color symmetric representation while the currents $J_{3}(x)$,
$J_{4}(x)$, $J_{8}(x)$, $J_{9}(x)$, $J_{10}(x)$, $J_{11\mu\nu}(x)$,
$J_{12\mu\nu}(x)$ belong to the $[\mathbf{\bar 3_c}]_{qc} \otimes
[\mathbf{3_c}]_{\bar{q}\bar c}$ color antisymmetric representation.
Throughout our calculation, we assume $m_u=m_d=0$. Hence the masses
of the isoscalar and isovector tetraquark states with the same heavy
flavor content are degenerate.

We study the two-point correlation functions induced by the above
scalar and tensor interpolating currents respectively
\begin{align}
\Pi(q^2)&=i\int d^4x \,e^{iq\cdot x}\,\langle 0|T [J(x)J^\dagger(0)]|0\rangle\, , \label{piscalar}\\
\Pi_{\mu\nu,\rho\sigma}(q^2)&=i\int d^4x\,e^{iq\cdot
x}\,\langle0|T[J_{\mu\nu}(x)J_{\rho\sigma}^{\dag}(0)]|0\rangle\,
,\label{pitensor}
\end{align}
where the currents $J(x)$ and $J_{\mu\nu}(x)$ can couple to the
corresponding hadronic states with the same quantum numbers
\begin{align}
\langle0|J|X\rangle&=f_{S}\,, \label{scalarcoupling}\\
\langle0|J_{\mu\nu}|X\rangle&=f_{T} \epsilon_{\mu\nu}+\cdots\, ,
\label{tensorcoupling}
\end{align}
in which $\epsilon_{\mu\nu}$ is the polarization tensor, $f_S$ and $f_T$
are the coupling constants. The polarization tensor
$\epsilon_{\mu\nu}$ in Eq.~\eqref{tensorcoupling} represents the
coupling to the spin-2 state. There also exist some other
structures (represented by ``$\cdots$") for spin-0 and spin-1
hadrons, which are omitted here. Accordingly, the correlation
function for the tensor current in Eq.~\eqref{pitensor} can be
written as
\begin{eqnarray}
\Pi_{\mu\nu,\rho\sigma}(q^2)=\frac{1}{2}(\eta_{\mu\rho}\eta_{\nu\sigma}+\eta_{\mu\sigma}\eta_{\nu\rho}
-\frac{2}{3}\eta_{\mu\nu}\eta_{\rho\sigma})\Pi(q^2)+\cdots\, ,
\label{equ:tensorPi}
\end{eqnarray}
where $\eta_{\mu\nu}=q_\mu q_\nu/q^2-g_{\mu\nu}$. At the hadronic
level, this invariant function can be described by the dispersion
relation
\begin{eqnarray}
\Pi(q^2)=\frac{(q^2)^N}{\pi}\int_{4m_c^2}^{\infty}\frac{\mbox{Im}\Pi(s)}{s^N(s-q^2-i\epsilon)}ds+\sum_{n=0}^{N-1}b_n(q^2)^n\,
, \label{Phenpi}
\end{eqnarray}
in which $b_n$ are unknown subtraction constants. The imaginary part
in the first term is defined as the spectral function and can be
written as a sum over $\delta$ functions
\begin{align}
\nonumber\rho(s)\equiv\mbox{Im}\Pi(s)/\pi&=\sum_n\delta(s-m_n^2)\langle0|J|n\rangle\langle n|J^{\dagger}|0\rangle \\
&=f_X^2\delta(s-m_X^2)+ \mbox{continuum}\, ,  \label{Phenrho}
\end{align}
where we adopt the single narrow pole plus continuum parametrization
in the second step.

Using the operator product expansion (OPE) method, the correlation
function can also be computed at the quark-gluonic level in
expression of various QCD condensates. One can then establish QCD
sum rules due to the quark-hadron duality that the correlation
functions obtained at the hadronic and quark-gluonic levels must
equal to each other. After performing the Borel transform, the QCD
sum rules read as functions of the continuum threshold $s_0$ and
Borel parameter $M_B^2$
\begin{eqnarray}
\mathcal{L}_{k}(s_0,
M_B^2)=f_X^2m_X^{2k}e^{-m_X^2/M_B^2}=\int_{4m_c^2}^{s_0}dse^{-s/M_B^2}\rho(s)s^k\,
. \label{sumrule}
\end{eqnarray}
The mass of the lowest-lying hadron state can be extracted as
\begin{eqnarray}
m_X(s_0, M_B^2)=\sqrt{\frac{\mathcal{L}_{1}(s_0,
M_B^2)}{\mathcal{L}_{0}(s_0, M_B^2)}}\, . \label{mass}
\end{eqnarray}

In this paper, the spectral density in Eq.~\eqref{sumrule} is
calculated up to dimension eight at the leading order of $\alpha_s$,
including the perturbative term and various non-perturbative
condensates. In Appendix~\ref{sec:rhos}, we list the expressions of
$\rho(s)$ for all interpolating currents in
Eqs.~\eqref{scalarcurrents}-\eqref{tensorcurrents}.

 \section{Numerical analysis}\label{Sec:NA}

In this section, we perform numerical analyses using the following
parameters of quark masses and various QCD
condensates~\cite{2016-Patrignani-p100001-100001,2001-Eidemuller-p203-210,
1999-Jamin-p300-303,
2002-Jamin-p237-243,2011-Khodjamirian-p94031-94031}:
\begin{eqnarray}
\nonumber &&m_c(m_c)=1.27\pm0.03\text{ GeV} \, , \non
&&m_b(m_b)=4.18^{+0.04}_{-0.03}\text{ GeV} \, , \non
&&\qq=-(0.23\pm0.03)^3\text{ GeV}^3 \, ,
\\ &&\qGqa=-M_0^2\qq\, ,
\non &&M_0^2=(0.8\pm0.2)\text{ GeV}^2 \, ,
\non&&\GGb=(0.48\pm0.14)\text{ GeV}^4 \, , \label{parameters}
\end{eqnarray}
in which the $\overline{MS}$ running heavy quark masses are adopted.
The QCD sum rules in Eq.~\eqref{sumrule} are functions of the
continuum threshold $s_0$ and Borel parameter $M_B^2$. The working
ranges for these two parameters will affect the numerical sum rule
analyses. The suitable working range (Borel window) of $M_B^2$ can
be determined by the requirement of the OPE convergence and the pole
contribution (PC). In our analyses, we use the following criteria to
obtain the Borel windows and optimal values for $s_0$:
\begin{enumerate}
\item Requiring the dominant non-perturbative contribution (quark condensate $\qq$) to be less than at least one
half of the perturbative term leads to the lower bound on the Borel
parameter. This ratio is adjusted as one third for the currents
$J_7(x)$ and $J_{10}(x)$ since the quark condensate $\qq$ and
quark-gluon mixed condensate $\qGqa$ give no contribution in OPEs
and thus the dimension six condensate $\qq^2$ is the dominant power
correction for these two channels.
\item The contribution of the dimension eight condensate $\qq\qGqa$ should be less than $5\%$. This requirement can
be usually satisfied under the first criterion except for the
$J_7(x)$ and $J_{10}(x)$.
\item We require the pole contribution to be larger than $10\%$ ($30\%$ for $J_8(x)$) to restrict the upper bound on the Borel parameter, in which the PC is defined as
\begin{eqnarray}
\mbox{PC}\equiv\frac{\mathcal{L}_{0}(s_0,
M_B^2)}{\mathcal{L}_{0}(\infty,
M_B^2)}=\frac{\int_{4m_c^2}^{s_0}dse^{-s/M_B^2}\rho(s)}{\int_{4m_c^2}^{\infty}dse^{-s/M_B^2}\rho(s)}\,
. \label{pc}
\end{eqnarray}
\item By minimizing the dependence of $m_X$ on $M_B^2$, we can determine the optimal value of $s_0$ in the Borel
window.
\end{enumerate}

\begin{center}
\begin{tabular}{lr}
\scalebox{0.8}{\includegraphics{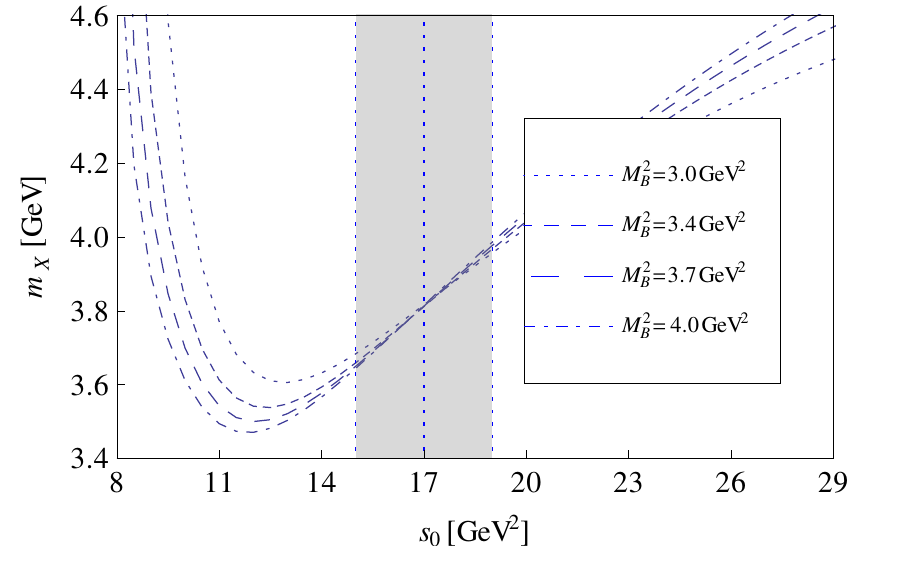}}&
\scalebox{0.8}{\includegraphics{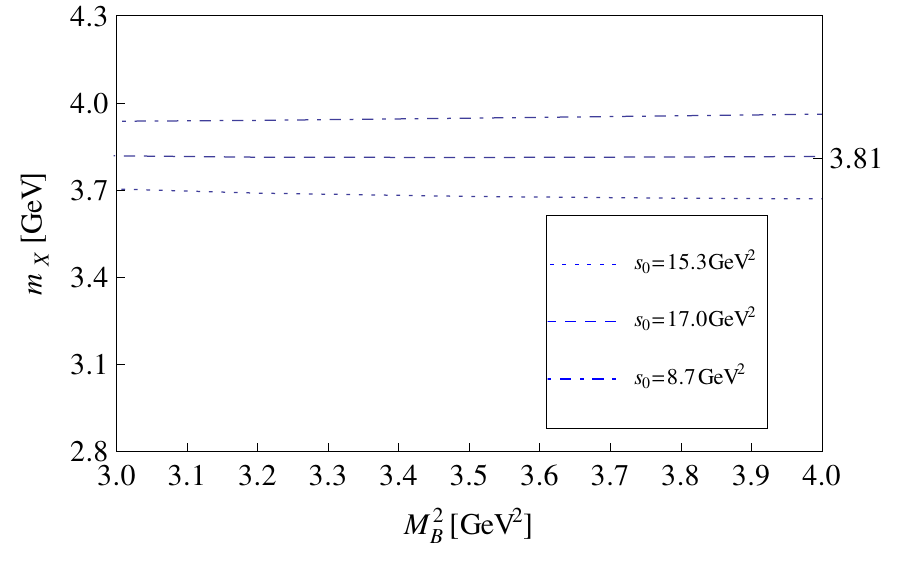}}
\end{tabular}
\figcaption{Variations of the $qc\bar q\bar c$ hadron mass $m_X$
with $s_0$ and $M_B^2$ for the $J^{PC}=0^{++}$ tetraquark using
current $J_4(x)$.} \label{0++4}
\end{center}

The advantage of these criteria is that the working ranges for $s_0$
and $M_B^2$ can be determined by the intrinsic  behavior of QCD sum
rules itself. To show the behavior of the mass sum rules, we plot
the variations of the extracted hadron mass with respect to $s_0$
and $M_B^2$ for the scalar current $J_4(x)$ in Fig.~\ref{0++4} as an
example. Applying the above criteria, the Borel window for $J_4(x)$
is determined to be $3.0\,{\rm GeV^2}\leq M_B^2\leq4.0\,{\rm GeV^2}$
with the optimal continuum threshold value $s_0=17.0$ GeV$^2$. One
may find from the left side of Fig.~\ref{0++4} that the curves of
$m_X$ with different value of $M_B^2$ intersect around $s_0=17.0$
GeV$^2$, where the variation of $m_X$ with $M_B^2$ is very weak.
Considering $10\%$ uncertainty of $s_0$, we can plot the Borel
curves in the above Borel window, as shown in the right side of
Fig~\ref{0++4}. These Borel curves are very stable with respect to
$M_B^2$ and thus we extract the hadron mass and coupling constant as
\begin{eqnarray}
m_{X,\, 0^{++}}&=&3.81\pm0.19~\text{GeV}\, , \\
f_{X,\, 0^{++}}&=&2.18\times 10^{-2}~\text{GeV}^5\, ,
\end{eqnarray}
which is in very good agreement with the experimental mass of the
$X^\ast(3860)$ state.

\begin{table}
\renewcommand{\arraystretch}{1.3}
\begin{tabular*}{12cm}{ccccccc}
\hlinewd{.8pt}
~$J^{PC}$   ~& ~Currents  ~& ~ $s_0(\mbox{GeV}^2)$ ~ & ~$M_B^2$(\mbox{GeV}$^2$)~ & ~$m_X$ \mbox{(GeV)} ~&~~ \mbox{PC} ~~& ~$f_X$ \mbox{($10^{-2}$GeV$^5$)} \\
\hline
$0^{++}$       & $J_1$                &  $20\pm2.0$               & $4.1 - 5.0$           & $4.17\pm0.20$   & $13.9\%$  & $3.15$\\
               & $J_2$                &  $15\pm1.5$               & $3.0 - 3.6$           & $3.56\pm0.17$   & $14.4\%$  & $2.14$\\
               & $J_3$                &  $16\pm1.6$               & $4.0 - 4.3$           & $3.72\pm0.17$   & $9.41\%$  & $1.10$\\
               & $J_4$                &  $17\pm1.7$               & $3.0 - 4.0$           & $3.81\pm0.19$   & $15.9\%$  & $2.18$\\
               & $J_7$                &  $15\pm1.5$               & $2.6 - 3.4$           & $3.58\pm0.18$   & $16.0\%$  & $3.77$\\
               & $J_9$                &  $19\pm1.9$               & $3.1 - 3.4$           & $3.93\pm0.19$   & $12.2\%$  & $1.42$\\
               & $J_{10}$             &  $18\pm1.8$               & $3.1 - 3.9$           & $3.90\pm0.16$   & $14.4\%$  & $4.99$\\
\hline
$2^{++}$       & $J_{11\mu\nu}$       &  $19\pm1.9$               & $4.2 - 4.8$           & $4.06\pm0.15$   & $12.8\%$  & $11.0$\\
               & $J_{13\mu\nu}$       &  $20\pm2.0$               & $4.2 - 5.1$           & $4.16\pm0.20$   & $14.3\%$  & $18.6$\\
\hlinewd{.8pt}
\end{tabular*}
\caption{Masses of the charmonium-like $qc\bar q\bar c$ tetraquark
states. The mass sum rules are unstable for the interpolating
currents $J_5(x)$, $J_{12\mu\nu}(x)$ and $J_{14\mu\nu}(x)$.
\label{qcqc}}
\end{table}
\begin{table}
\renewcommand{\arraystretch}{1.3}
\begin{tabular*}{12cm}{ccccccc}
\hlinewd{.8pt}
~$J^{PC}$   ~& ~Currents  ~& ~ $s_0(\mbox{GeV}^2)$ ~ & ~$M_B^2$(\mbox{GeV}$^2$)~ & ~$m_X$ \mbox{(GeV)} ~&~~ \mbox{PC} ~~& ~$f_X$ \mbox{($10^{-2}$GeV$^5$)} \\
\hline
$0^{++}$       & $J_1$                &  $20\pm2.0$               & $3.7 - 4.9$           & $4.18\pm0.19$   & $14.8\%$  & $2.97$\\
               & $J_2$                &  $15\pm1.5$               & $2.7 - 3.5$           & $3.57\pm0.15$   & $15.5\%$  & $1.94$\\
               & $J_3$                &  $16\pm1.6$               & $3.7 - 4.0$           & $3.73\pm0.17$   & $10.6\%$  & $1.00$\\
               & $J_4$                &  $17\pm1.7$               & $2.7 - 3.9$           & $3.83\pm0.19$   & $17.0\%$  & $2.13$\\
               & $J_7$                &  $16\pm1.6$               & $2.4 - 3.4$           & $3.61\pm0.15$   & $19.1\%$  & $4.39$\\
               & $J_9$                &  $18\pm1.8$               & $2.8 - 3.1$           & $3.86\pm0.15$   & $13.4\%$  & $1.19$\\
               & $J_{10}$             &  $18\pm1.8$               & $2.8 - 3.9$           & $3.92\pm0.17$   & $16.1\%$  & $4.86$\\
\hline
$2^{++}$       & $J_{11\mu\nu}$       &  $19\pm1.9$               & $3.8 - 4.7$           & $4.07\pm0.20$   & $14.2\%$  & $10.3$\\
               & $J_{13\mu\nu}$       &  $20\pm2.0$               & $3.8 - 5.0$           & $4.17\pm0.19$   & $15.0\%$  & $17.5$\\
\hlinewd{.8pt}
\end{tabular*}
\caption{Masses of the charmonium-like $sc\bar s\bar c$ tetraquark
states.
 \label{scsc}}
\end{table}

Performing similar analyses, we study the mass sum rules for all
interpolating currents in Eq.~\eqref{scalarcurrents}. We study the
properties of the spectral densities in Fig.~\ref{rho}. The spectral
density for $J_4(x)$ becomes positive in the region $s>7.5$ GeV$^2$.
However, the behavior for the spectral density for $J_8(x)$ is more
complicated, as shown in Fig.~\ref{rho}, which becomes positive only
for $s>18.5$ GeV$^2$. Such a spectral density is unphysical and can
not be used to make a reliable mass prediction. The situations are
similar for the currents $J_5(x)$ and $J_6(x)$. We shall not make
mass predictions using these currents. For the other interpolating
currents, we perform the numerical analyses and collect the
numerical results in Table~\ref{qcqc}. The errors of $m_X$ come from
the uncertainties of charm quark mass, various condensates and the
continuum threshold $s_0$, in which the uncertainties from $s_0$ and
quark condensate are the dominant error sources. As shown in
Table~\ref{qcqc}, the masses extracted from $J_4(x), J_9(x)$ and
$J_{10}(x)$ are very close to the mass of $X^\ast(3860)$, which
implies that these currents may well couple to this state and
suggests a possible tetraquark interpretation for $X^\ast(3860)$.

\begin{center}
\begin{tabular}{lr}
\scalebox{0.8}{\includegraphics{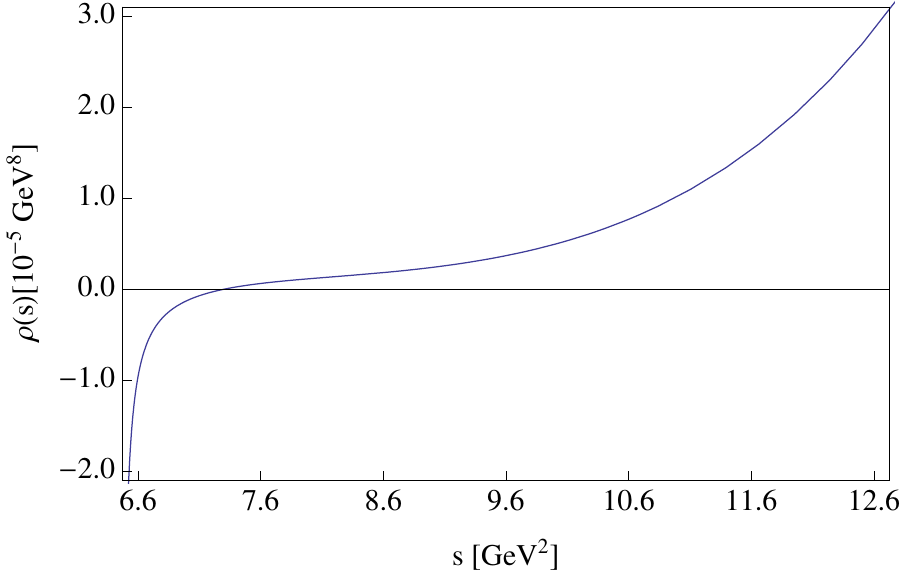}}&
\scalebox{0.8}{\includegraphics{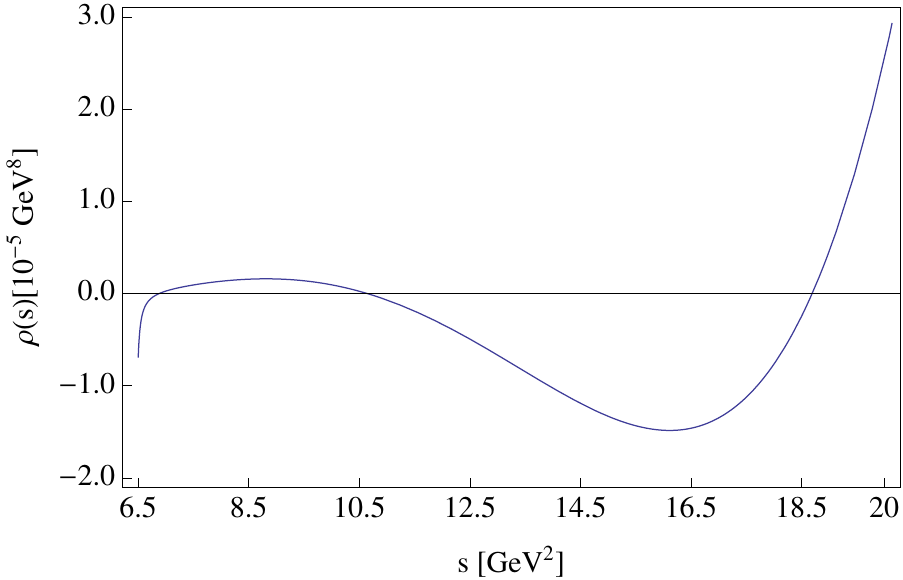}}
\end{tabular}
\figcaption{Property of the spectral density for the interpolating currents $J_4(x)$ (left) and $J_8(x)$ (right) with $J^{PC}=0^{++}$.} \label{rho}
\end{center}

For the tensor current $J_{11\mu\nu}$ with $J^{PC}=2^{++}$, we show
the variations of $m_X$ with $s_0$ and $M_B^2$ in
Fig.~\ref{fig2++11} and extract the mass and coupling constant as
\begin{eqnarray}
m_{X,\, 2^{++}}&=&4.06\pm0.15~\text{GeV}\, , \\
f_{X,\, 2^{++}}&=&0.11~\text{GeV}^5\, ,
\end{eqnarray}
which is a bit higher than the mass of $X^\ast(3860)$, but is still
consistent with the experiment result within errors. Similarly, we
can also study the hidden-charm $sc\bar s\bar c$ tetraquark systems
in the same channels. Using the spectral densities in
Appendix~\ref{sec:rhos}, we can make the replacement $\qq\to\ss$ and
$\qGqa\to\sGsa$. After performing similar numerical analyses, we
collect the numerical results for the $sc\bar s\bar c$ tetraquark
states in Table~\ref{scsc}. The masses for these $sc\bar s\bar c$
tetraquarks are almost degenerate with the $sc\bar s\bar c$ states
with the same current.

\begin{center}
\begin{tabular}{lr}
\scalebox{0.8}{\includegraphics{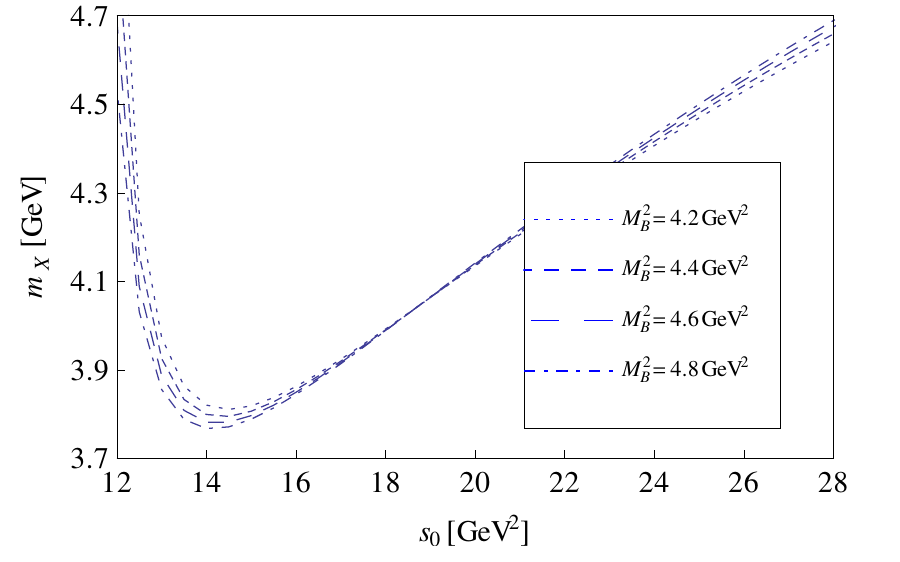}}&
\scalebox{0.8}{\includegraphics{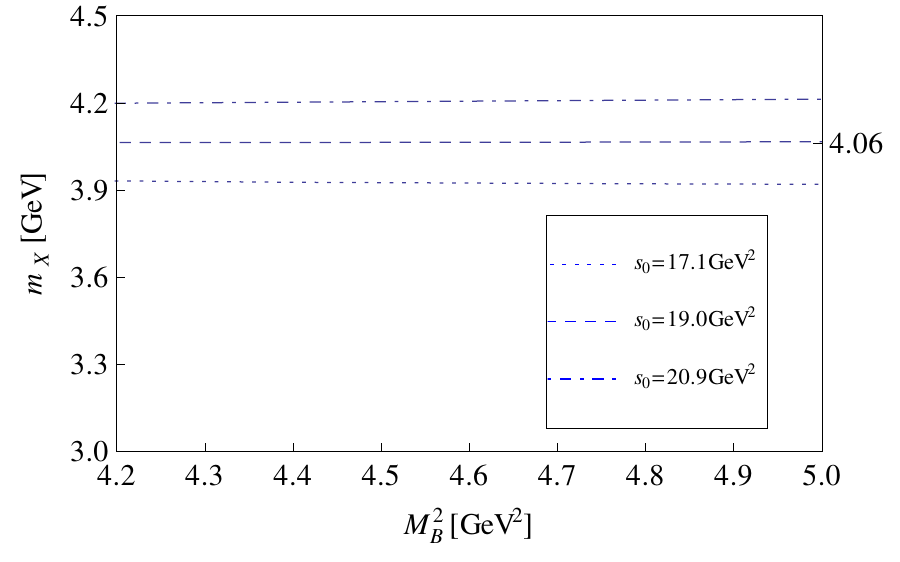}}
\end{tabular}
\figcaption{Variations of the $qc\bar q\bar c$ hadron mass $m_X$
with $s_0$ and $M_B^2$ for the $J^{PC}=2^{++}$ tetraquark using
current $J_{11\mu\nu}(x)$.} \label{fig2++11}
\end{center}

With the heavy quark symmetry, we can similarly study the
hidden-bottom $qb\bar q\bar b$ and $sb\bar s\bar b$ tetraquark
states with $J^{PC}=0^{++}$ and $2^{++}$. Replacing $m_c\to m_b$ in
the expressions of $\rho(s)$, we perform QCD sum rule analyses
collect the numerical results for the hidden-bottom $qb\bar q\bar b$
and $sb\bar s\bar b$  tetraquarks in Table~\ref{qbqb}. One notes
that the pole contributions and coupling constants for the
hidden-bottom tetraquark systems are much higher than those in the
hidden-charm systems. The masses for these hidden-bottom tetraquarks
are around $9.7-10.2$ GeV.

\begin{table}
\renewcommand{\arraystretch}{1.3}
\begin{tabular*}{12cm}{ccccccc}
\hlinewd{.8pt}
~$J^{PC}$   ~& ~Currents  ~& ~ $s_0(\mbox{GeV}^2)$ ~ & ~$M_B^2$(\mbox{GeV}$^2$)~ & ~$m_X$ \mbox{(GeV)} ~&~~ \mbox{PC} ~~& ~$f_X$ \mbox{($10^{-1}$GeV$^5$)} \\
\hline
$0^{++}$       & $J_1$                &  $108\pm5$               & $9.3 - 9.9$           & $10.00\pm0.21$   & $22.0\%$  & $1.51$\\
               & $J_2$                &  $103\pm5$               & $7.4 - 8.8$           & $9.71\pm0.19$   & $26.6\%$  & $1.82$\\
               & $J_3$                &  $104\pm5$               & $9.2 - 9.5$           & $9.83\pm0.20$   & $19.5\%$  & $0.82$\\
               & $J_4$                &  $106\pm5$               & $7.6 - 9.2$           & $9.87\pm0.20$   & $26.6\%$  & $1.59$\\
               & $J_7$                &  $105\pm5$               & $7.6 - 8.4$           & $9.80\pm0.22$   & $23.8\%$  & $4.03$\\
               & $J_9$                &  $112\pm5$               & $8.2 - 8.5$           & $10.18\pm0.22$   & $20.4\%$  & $1.56$\\
               & $J_{10}$            &  $108\pm5$               & $8.0 - 8.9$           & $9.96\pm0.21$   & $23.6\%$  & $3.65$\\
\hline
$2^{++}$       & $J_{11\mu\nu}$       &  $108\pm5$            & $9.5 - 10.1$           & $10.00\pm0.21$   & $21.8\%$  & $6.52$\\
               & $J_{13\mu\nu}$       &  $109\pm5$               & $9.5 - 10.2$           & $10.05\pm0.22$   & $22.7\%$  & $9.88$\\
\hlinewd{.8pt}
\end{tabular*}
\caption{Masses of the bottomonium-like $qb\bar q\bar b$ tetraquark
states. \label{qbqb}}
\end{table}

\begin{table}
\renewcommand{\arraystretch}{1.3}
\begin{tabular*}{12cm}{ccccccc}
\hlinewd{.8pt}
~$J^{PC}$   ~& ~Currents  ~& ~ $s_0(\mbox{GeV}^2)$ ~ & ~$M_B^2$(\mbox{GeV}$^2$)~ & ~$m_X$ \mbox{(GeV)} ~&~~ \mbox{PC} ~~& ~$f_X$ \mbox{($10^{-1}$GeV$^5$)} \\
\hline
$0^{++}$       & $J_1$                &  $108\pm5$               & $8.5 - 9.6$           & $10.01\pm0.21$   & $24.5\%$  & $1.42$\\
               & $J_2$                &  $103\pm5$               & $7.0 - 8.6$           & $9.72\pm0.19$   & $28.3\%$  & $1.70$\\
               & $J_3$                &  $104\pm5$               & $8.5 - 9.0$           & $9.84\pm0.20$   & $21.8\%$  & $0.76$\\
               & $J_4$                &  $106\pm5$               & $7.2 - 9.0$           & $9.88\pm0.19$   & $27.4\%$  & $1.51$\\
               & $J_7$                &  $105\pm5$               & $7.2 - 8.2$           & $9.82\pm0.20$   & $25.1\%$  & $3.95$\\
               & $J_9$                &  $111\pm5$               & $7.7 - 8.3$           & $10.15\pm0.21$   & $22.2\%$  & $1.51$\\
               & $J_{10}$            &  $108\pm5$               & $7.4 - 8.7$           & $9.97\pm0.19$   & $26.1\%$  & $3.60$\\
\hline
$2^{++}$       & $J_{11\mu\nu}$       &  $108\pm5$            & $8.7 - 9.8$           & $10.01\pm0.20$   & $24.1\%$  & $6.14$\\
               & $J_{13\mu\nu}$       &  $110\pm5$               & $8.7 - 10.2$           & $10.09\pm0.21$   & $25.4\%$  & $9.97$\\
\hlinewd{.8pt}
\end{tabular*}
\caption{Masses of the bottomonium-like $sb\bar s\bar b$ tetraquark
states. \label{sbsb}}
\end{table}

 \section{Discussion and Summary}\label{Sec:sum}
In this work, we have studied the hidden-charm/bottom $qc\bar q\bar
c$, $sc\bar s\bar c$ and $qb\bar q\bar b$, $sb\bar s\bar b$
tetraquark systems in the method of QCD sum rules. We have
constructed the interpolating tetraquark currents with
$J^{PC}=0^{++}$ and $2^{++}$ in a systematical way and calculated
their correlation functions and spectral densities at the leading
order on $\alpha_s$. The mass spectra for these scalar and tensor
tetraquark states are predicted. Since the quantum numbers for the
$X(4160)$, $X(3915)$, $X(4350)$ and $X^\ast(3860)$ can be
$J^{PC}=0^{++}$ or $2^{++}$, we can compare the experimental
results for these resonances with the tetraquark mass spectra listed
in Tables~\ref{qcqc}-\ref{scsc}.

\begin{center}
\begin{tabular}{c}
\scalebox{0.7}{\includegraphics{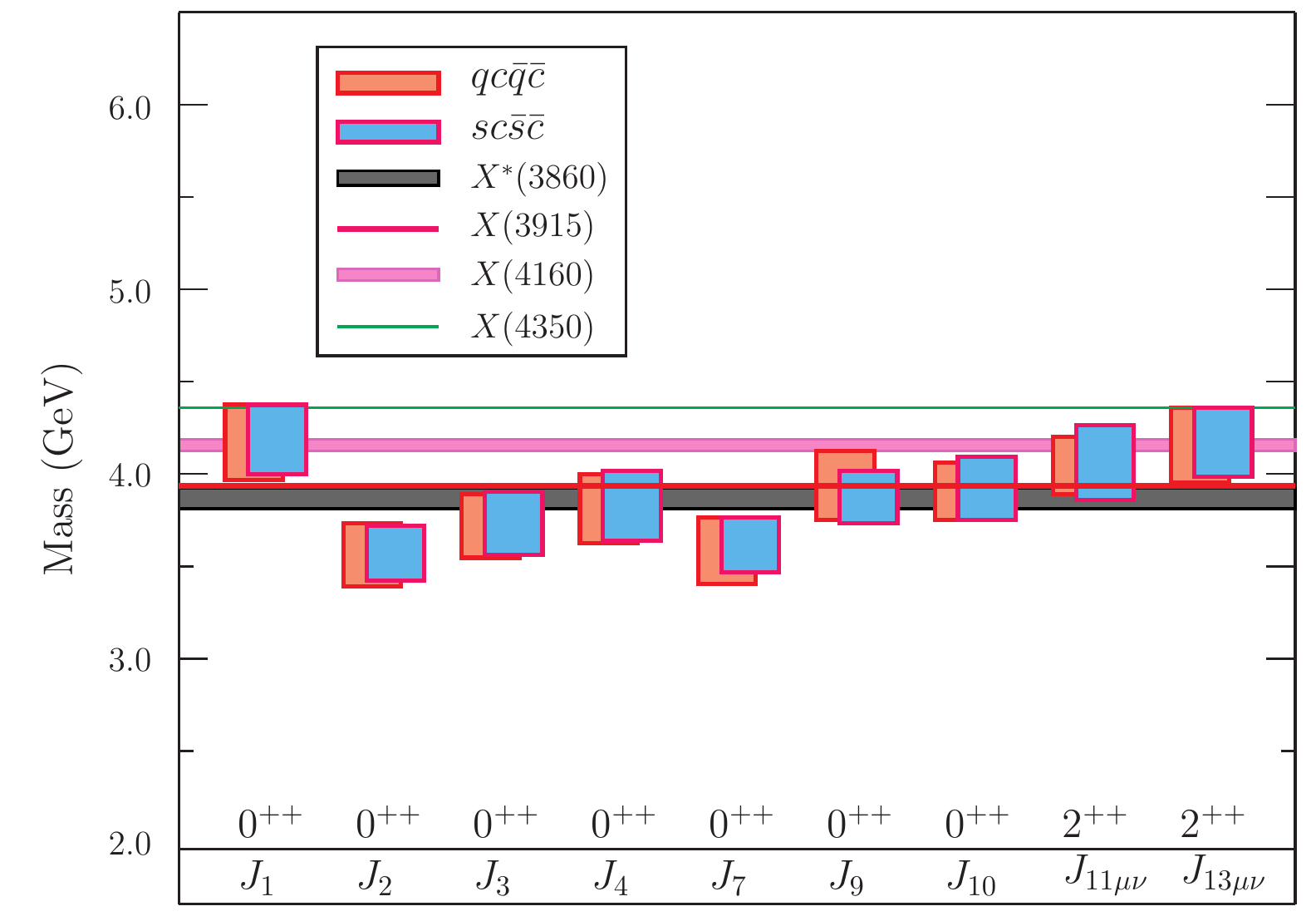}}
\end{tabular}
\figcaption{Mass spectra for the hidden-charm $qc\bar q\bar c$ and
$sc\bar s\bar c$ tetraquark states with $J^{PC}=0^{++}$ and
$2^{++}$. The vertical sizes of the rectangles represent the
uncertainties of the experimental hadron masses and our
calculations.} \label{Spectra}
\end{center}

In Fig.~\ref{Spectra}, we show the mass spectra for the hidden-charm
$qc\bar q\bar c$ and $sc\bar s\bar c$ tetraquark states labeled by
the interpolating current and $J^{PC}$ quantum numbers. To compare
these mass spectra with the masses of the $X(4160)$, $X(3915)$,
$X(4350)$ and $X^\ast(3860)$, we also show their experimental mass
values with uncertainties in Fig.~\ref{Spectra}.

For the hidden-charm $qc\bar q\bar c$ systems, the currents $J_4(x)$,
$J_9(x)$ and $J_{10}(x)$ are composed of color antisymmetric.
The isovector and isoscalar tetraquark masses
extracted from these three currents are about $3.8-3.9$ GeV, which is
consistent with the mass of the $X^\ast(3860)$ state, as shown in
Fig.~\ref{Spectra}. However, the isoscalar tetraquark currents
composed of two S-wave diquarks may also couple to a conventional
charmonium, especially the radially excited charmonium. For example,
the light tetraquark currents may also couple to the conventional
non-exotica physical states \cite{2005-Jaffe-p1-45}. In other words,
$X^\ast(3860)$ may be either an isoscalar tetraquark state or
$\chi_{c0}(2P)$.

In contrast, the masses for the tensor charmonium-like tetraquarks
are about $4.06-4.16$ GeV, which is a bit higher than that of
$X^\ast(3860)$ but with a small overlap within errors. On the other
hand, our results prefer the $J^{PC}=0^{++}$ assignment for the
$X^\ast(3860)$ over the $2^{++}$ assignment, which is also in
agreement with the Belle experiment~\cite{2017-Chilikin-p112003-112003}.
Nonetheless, the $2^{++}$ possibility is still not excluded as shown
in Fig.~\ref{Spectra}.

Using the currents $J_1(x)$ with $J^{PC}=0^{++}$ and
$J_{13\mu\nu}(x)$ with $J^{PC}=2^{++}$, we extract the hadron masses
for the scalar and tensor $qc\bar q\bar c$ tetraquarks around
$4.1-4.2$ GeV. These results are in good agreement with the mass of
the $X(4160)$ state, which implies the tetraquark interpretation for
this resonance. For the $X(3915)$, our results favor the $0^{++}$
$qc\bar q\bar c$ or $sc\bar s\bar c$ tetraquark assignment over the
tensor assignment. From Fig.~\ref{Spectra}, the $qc\bar q\bar c$ and
$sc\bar s\bar c$ tetraquarks are almost degenerate for the same
interpolating current and quantum numbers. Our results do not
support the $X(4350)$ to be a $sc\bar s\bar c$ tetraquark with
$J^{PC}=0^{++}$ or $2^{++}$.
We have also predicted the mass spectra of the hidden-bottom $qb\bar
q\bar b$ and $sb\bar s\bar b$ tetraquarks with $J^{PC}=0^{++}$ and
$2^{++}$. The masses for these hidden-bottom tetraquarks are
obtained around $9.7-10.2$ GeV.
These mass predictions may be useful for
understanding the tetraquark spectroscopy and searching for such
states at the facilities such as LHCb and BelleII in the future.

\section*{Acknowledgments}

This project is supported by the Natural Sciences and Engineering Research Council of
Canada (NSERC) and the National Natural Science Foundation of China under Grants
No. 11475015, No. 11375024, No. 11222547, No. 11175073, No. 11575008, and No. 11621131001;
the 973 program; the Ministry of Education of China (SRFDP under Grant No. 20120211110002 and the Fundamental Research
Funds for the Central Universities); the National Program for Support of Top-notch Youth Professionals.



\appendix

\section{The spectral densities}\label{sec:rhos}
In this appendix, we list the expressions of the spectral densities
for the interpolating currents listed in Eqs.
\eqref{scalarcurrents}-\eqref{tensorcurrents} as the following. The
spectral densities are calculated by including the perturbative
term, quark condensate $\qq$, gluon condensate $\GGb$, quark-gluon
mixed condensate $\qGqa$, four-quark condensate $\qq^2$ and the
dimension eight condensate $\qq\qGqa$
\begin{eqnarray}
\rho_i(s)=\rho_i^{pert}(s)+\rho_i^{\qq}(s)+\rho_i^{\GGa}(s)+\rho_i^{\qGqb}(s)+\rho_i^{\qq^2}(s)+\rho_i^{\qq\qGqb}(s)\,
,
\end{eqnarray}
in which the subscript ``$i$" denotes the interpolating current
number.

\begin{itemize}
\item For the current $J_1$ with $J^{PC}=0^{++}$:
{\allowdisplaybreaks
\begin{eqnarray}
\nonumber
\rho^{pert}_1(s)&=&\frac{1}{256\pi^6}\dab\frac{(1-\alpha-\beta)^2\left[m_c^2(\alpha+\beta)-3\alpha\beta
s\right]\f(s)^3}{\alpha^3\beta^3}\, , \non
\rho^{\qq}_1(s)&=&-\frac{m_c\qq}{4\pi^4}\dab\frac{(1-\alpha-\beta)\f(s)\left[m_c^2(\alpha+\beta)-2\alpha\beta
s\right]}{\alpha^2\beta}\, , \non
\rho^{\GGa}_1(s)&=&\frac{\GGb}{256\pi^6}\dab\Bigg\{\frac{m_c^2(1-\alpha-\beta)^2\left[2m_c^2(\alpha+\beta)-3\alpha\beta
s\right]}{3\beta^3} \non&&
-\frac{(1-\alpha-\beta)\left[m_c^2(\alpha+\beta)-2\alpha\beta
s\right]\f(s)}{2\alpha\beta^2}\Bigg\}\, , \non
\rho^{\qGqb}_1(s)&=&\frac{m_c\qGqa}{32\pi^4}\dab
\frac{(1+\alpha-\beta)\left[2m_c^2(\alpha+\beta)-3\alpha\beta
s\right]}{\alpha^2}\, , \non
\rho^{\qq^2}_1(s)&=&\frac{m_c^2\qq^2}{6\pi^2}\sqrt{1-4m_c^2/s}\, ,
\\
\rho^{\qq\qGqb}_1(s)&=&\frac{m_c^2\qq\qGqa}{24\pi^2}\int_0^1d\alpha
\Bigg\{\frac{2m_c^2}{\alpha^2}\delta'\left(s-\tilde{m}^2_c\right)+
\frac{1}{\alpha}\delta\left(s-\tilde{m}^2_c\right)\Bigg\}\, ,
\label{SD0++1}
\end{eqnarray}
} where the integral limits are
\begin{eqnarray}
\alpha_{max}=\frac{1+\sqrt{1-4m_c^2/s}}{2}\, , \, \,
\alpha_{min}=\frac{1-\sqrt{1-4m_c^2/s}}{2}\, , \, \,
\beta_{max}=1-\alpha\, , \, \, \beta_{min}=\frac{\alpha
m_c^2}{\alpha s-m_c^2}\, .
\end{eqnarray}

\item For the current $J_2$ with $J^{PC}=0^{++}$:
{\allowdisplaybreaks
\begin{eqnarray}
\nonumber
\rho^{pert}_2(s)&=&\frac{1}{64\pi^6}\dab\frac{(1-\alpha-\beta)^2\left[m_c^2(\alpha+\beta)-3\alpha\beta
s\right]\f(s)^3}{\alpha^3\beta^3}\, , \non
\rho^{\qq}_2(s)&=&-\frac{m_c\qq}{2\pi^4}\dab\frac{(1-\alpha-\beta)\f(s)\left[m_c^2(\alpha+\beta)-2\alpha\beta
s\right]}{\alpha^2\beta}\, , \non
\rho^{\GGa}_2(s)&=&\frac{\GGb}{64\pi^6}\dab\Bigg\{\frac{m_c^2(1-\alpha-\beta)^2\left[2m_c^2(\alpha+\beta)-3\alpha\beta
s\right]}{3\beta^3} \non&&
+\frac{5\left[m_c^2(\alpha+\beta)-2\alpha\beta
s\right]\f(s)}{16}\left[\frac{(1-\alpha-\beta)^2}{2\alpha^2\beta^2}+\frac{(2-2\alpha-\beta)}{\alpha\beta^2}\right]\Bigg\}\,
, \non \rho^{\qGqb}_2(s)&=&-\frac{m_c\qGqa}{64\pi^4}\dab
\frac{(5-13\beta)\left[2m_c^2(\alpha+\beta)-3\alpha\beta
s\right]}{\alpha\beta}\, , \non
\rho^{\qq^2}_2(s)&=&\frac{2m_c^2\qq^2}{3\pi^2}\sqrt{1-4m_c^2/s}\, ,
\\
\rho^{\qq\qGqb}_2(s)&=&\frac{m_c^2\qq\qGqa}{24\pi^2}\int_0^1d\alpha
\Bigg\{\frac{8m_c^2}{\alpha^2}\delta'\left(s-\tilde{m}^2_c\right)-
\frac{5}{\alpha}\delta\left(s-\tilde{m}^2_c\right)\Bigg\}\, .
\label{SD0++2}
\end{eqnarray}
}

\item For the current $J_3$ with $J^{PC}=0^{++}$:
{\allowdisplaybreaks
\begin{eqnarray}
\nonumber
\rho^{pert}_3(s)&=&\frac{1}{512\pi^6}\dab\frac{(1-\alpha-\beta)^2\left[m_c^2(\alpha+\beta)-3\alpha\beta
s\right]\f(s)^3}{\alpha^3\beta^3}\, , \non
\rho^{\qq}_3(s)&=&-\frac{m_c\qq}{8\pi^4}\dab\frac{(1-\alpha-\beta)\f(s)\left[m_c^2(\alpha+\beta)-2\alpha\beta
s\right]}{\alpha^2\beta}\, , \non
\rho^{\GGa}_3(s)&=&\frac{\GGb}{512\pi^6}\dab\Bigg\{\frac{m_c^2(1-\alpha-\beta)^2\left[2m_c^2(\alpha+\beta)-3\alpha\beta
s\right]}{3\beta^3} \non&&
+\frac{(1-\alpha-\beta)\left[m_c^2(\alpha+\beta)-2\alpha\beta
s\right]\f(s)}{\alpha\beta^2}\Bigg\}\, , \non
\rho^{\qGqb}_3(s)&=&-\frac{m_c\qGqa}{32\pi^4}\dab
\frac{(1-2\alpha-\beta)\left[2m_c^2(\alpha+\beta)-3\alpha\beta
s\right]}{\alpha^2}\, , \non
\rho^{\qq^2}_3(s)&=&\frac{m_c^2\qq^2}{12\pi^2}\sqrt{1-4m_c^2/s}\, ,
\\
\rho^{\qq\qGqb}_3(s)&=&\frac{m_c^2\qq\qGqa}{24\pi^2}\int_0^1d\alpha
\Bigg\{\frac{m_c^2}{\alpha^2}\delta'\left(s-\tilde{m}^2_c\right)-
\frac{1}{\alpha}\delta\left(s-\tilde{m}^2_c\right)\Bigg\}\, .
\label{SD0++3}
\end{eqnarray}
}

\item For the current $J_4$ with $J^{PC}=0^{++}$:
{\allowdisplaybreaks
\begin{eqnarray}
\nonumber
\rho^{pert}_4(s)&=&\frac{1}{128\pi^6}\dab\frac{(1-\alpha-\beta)^2\left[m_c^2(\alpha+\beta)-3\alpha\beta
s\right]\f(s)^3}{\alpha^3\beta^3}\, , \non
\rho^{\qq}_4(s)&=&-\frac{m_c\qq}{4\pi^4}\dab\frac{(1-\alpha-\beta)\f(s)\left[m_c^2(\alpha+\beta)-2\alpha\beta
s\right]}{\alpha^2\beta}\, , \non
\rho^{\GGa}_4(s)&=&\frac{\GGb}{128\pi^6}\dab\Bigg\{\frac{m_c^2(1-\alpha-\beta)^2\left[2m_c^2(\alpha+\beta)-3\alpha\beta
s\right]}{3\beta^3} \non&&
+\frac{\left[m_c^2(\alpha+\beta)-2\alpha\beta
s\right]\f(s)}{8}\left[\frac{(1-\alpha-\beta)^2}{2\alpha^2\beta^2}+\frac{(2-2\alpha-\beta)}{\alpha\beta^2}\right]\Bigg\}\,
, \non \rho^{\qGqb}_4(s)&=&-\frac{m_c\qGqa}{64\pi^4}\dab
\frac{(1-5\beta)\left[2m_c^2(\alpha+\beta)-3\alpha\beta
s\right]}{\alpha\beta}\, , \non
\rho^{\qq^2}_4(s)&=&\frac{m_c^2\qq^2}{3\pi^2}\sqrt{1-4m_c^2/s}\, ,
\\
\rho^{\qq\qGqb}_4(s)&=&\frac{m_c^2\qq\qGqa}{24\pi^2}\int_0^1d\alpha
\Bigg\{\frac{4m_c^2}{\alpha^2}\delta'\left(s-\tilde{m}^2_c\right)-
\frac{1}{\alpha}\delta\left(s-\tilde{m}^2_c\right)\Bigg\}\, .
\label{SD0++4}
\end{eqnarray}
}

\item For the current $J_5$ with $J^{PC}=0^{++}$:
{\allowdisplaybreaks
\begin{eqnarray}
\nonumber
\rho^{pert}_5(s)&=&\frac{1}{256\pi^6}\dab\frac{(1-\alpha-\beta)^2\left[m_c^2(\alpha+\beta)-3\alpha\beta
s\right]\f(s)^3}{\alpha^3\beta^3}\, , \non
\rho^{\qq}_5(s)&=&\frac{m_c\qq}{4\pi^4}\dab\frac{(1-\alpha-\beta)\f(s)\left[m_c^2(\alpha+\beta)-2\alpha\beta
s\right]}{\alpha^2\beta}\, , \non
\rho^{\GGa}_5(s)&=&\frac{\GGb}{256\pi^6}\dab\Bigg\{\frac{m_c^2(1-\alpha-\beta)^2\left[2m_c^2(\alpha+\beta)-3\alpha\beta
s\right]}{3\beta^3} \non&&
-\frac{(1-\alpha-\beta)\left[m_c^2(\alpha+\beta)-2\alpha\beta
s\right]\f(s)}{2\alpha\beta^2}\Bigg\}\, , \non
\rho^{\qGqb}_5(s)&=&-\frac{m_c\qGqa}{32\pi^4}\dab
\frac{(1+\alpha-\beta)\left[2m_c^2(\alpha+\beta)-3\alpha\beta
s\right]}{\alpha^2}\, , \non
\rho^{\qq^2}_5(s)&=&\frac{m_c^2\qq^2}{6\pi^2}\sqrt{1-4m_c^2/s}\, ,
\\
\rho^{\qq\qGqb}_5(s)&=&\frac{m_c^2\qq\qGqa}{24\pi^2}\int_0^1d\alpha
\Bigg\{\frac{2m_c^2}{\alpha^2}\delta'\left(s-\tilde{m}^2_c\right)+
\frac{1}{\alpha}\delta\left(s-\tilde{m}^2_c\right)\Bigg\}\, .
\label{SD0++5}
\end{eqnarray}
}

\item For the current $J_6$ with $J^{PC}=0^{++}$:
{\allowdisplaybreaks
\begin{eqnarray}
\nonumber
\rho^{pert}_6(s)&=&\frac{1}{64\pi^6}\dab\frac{(1-\alpha-\beta)^2\left[m_c^2(\alpha+\beta)-3\alpha\beta
s\right]\f(s)^3}{\alpha^3\beta^3}\, , \non
\rho^{\qq}_6(s)&=&\frac{m_c\qq}{2\pi^4}\dab\frac{(1-\alpha-\beta)\f(s)\left[m_c^2(\alpha+\beta)-2\alpha\beta
s\right]}{\alpha^2\beta}\, , \non
\rho^{\GGa}_6(s)&=&\frac{\GGb}{64\pi^6}\dab\Bigg\{\frac{m_c^2(1-\alpha-\beta)^2\left[2m_c^2(\alpha+\beta)-3\alpha\beta
s\right]}{3\beta^3} \non&&
+\frac{5\left[m_c^2(\alpha+\beta)-2\alpha\beta
s\right]\f(s)}{16}\left[\frac{(1-\alpha-\beta)^2}{2\alpha^2\beta^2}+\frac{(2-2\alpha-\beta)}{\alpha\beta^2}\right]\Bigg\}\,
, \non \rho^{\qGqb}_6(s)&=&\frac{m_c\qGqa}{64\pi^4}\dab
\frac{(5-13\beta)\left[2m_c^2(\alpha+\beta)-3\alpha\beta
s\right]}{\alpha\beta}\, , \non
\rho^{\qq^2}_6(s)&=&\frac{2m_c^2\qq^2}{3\pi^2}\sqrt{1-4m_c^2/s}\, ,
\\
\rho^{\qq\qGqb}_6(s)&=&\frac{m_c^2\qq\qGqa}{24\pi^2}\int_0^1d\alpha
\Bigg\{\frac{8m_c^2}{\alpha^2}\delta'\left(s-\tilde{m}^2_c\right)-
\frac{5}{\alpha}\delta\left(s-\tilde{m}^2_c\right)\Bigg\}\, .
\label{SD0++6}
\end{eqnarray}
}

\item For the current $J_7$ with $J^{PC}=0^{++}$:
{\allowdisplaybreaks
\begin{eqnarray}
\nonumber
\rho^{pert}_7(s)&=&\frac{3}{32\pi^6}\dab\frac{(1-\alpha-\beta)^2\left[m_c^2(\alpha+\beta)-3\alpha\beta
s\right]\f(s)^3}{\alpha^3\beta^3}\, , \non
\rho^{\qq}_7(s)&=&\rho^{\qGqb}_7(s)=0\, , \non
\rho^{\GGa}_7(s)&=&\frac{\GGb}{32\pi^6}\dab\Bigg\{\frac{m_c^2(1-\alpha-\beta)^2\left[2m_c^2(\alpha+\beta)-3\alpha\beta
s\right]}{\beta^3} \non&&
+\frac{\left[m_c^2(\alpha+\beta)-2\alpha\beta
s\right]\f(s)}{4}\left[\frac{5(1-\alpha-\beta)^2}{2\alpha^2\beta^2}+\frac{(12-12\alpha-7\beta)}{\alpha\beta^2}\right]\Bigg\}\,
, \non
\rho^{\qq^2}_7(s)&=&\frac{4m_c^2\qq^2}{\pi^2}\sqrt{1-4m_c^2/s}\, ,
\\
\rho^{\qq\qGqb}_7(s)&=&\frac{2m_c^2\qq\qGqa}{\pi^2}\int_0^1d\alpha
\Bigg\{\frac{m_c^2}{\alpha^2}\delta'\left(s-\tilde{m}^2_c\right)-
\frac{1}{\alpha}\delta\left(s-\tilde{m}^2_c\right)\Bigg\}\, .
\label{SD0++7}
\end{eqnarray}
}

\item For the current $J_8$ with $J^{PC}=0^{++}$:
{\allowdisplaybreaks
\begin{eqnarray}
\nonumber
\rho^{pert}_8(s)&=&\frac{1}{512\pi^6}\dab\frac{(1-\alpha-\beta)^2\left[m_c^2(\alpha+\beta)-3\alpha\beta
s\right]\f(s)^3}{\alpha^3\beta^3}\, , \non
\rho^{\qq}_8(s)&=&\frac{m_c\qq}{8\pi^4}\dab\frac{(1-\alpha-\beta)\f(s)\left[m_c^2(\alpha+\beta)-2\alpha\beta
s\right]}{\alpha^2\beta}\, , \non
\rho^{\GGa}_8(s)&=&\frac{\GGb}{512\pi^6}\dab\Bigg\{\frac{m_c^2(1-\alpha-\beta)^2\left[2m_c^2(\alpha+\beta)-3\alpha\beta
s\right]}{3\beta^3} \non&&
+\frac{(1-\alpha-\beta)\left[m_c^2(\alpha+\beta)-2\alpha\beta
s\right]\f(s)}{\alpha\beta^2}\Bigg\}\, , \non
\rho^{\qGqb}_8(s)&=&\frac{m_c\qGqa}{32\pi^4}\dab
\frac{(1-2\alpha-\beta)\left[2m_c^2(\alpha+\beta)-3\alpha\beta
s\right]}{\alpha^2}\, , \non
\rho^{\qq^2}_8(s)&=&\frac{m_c^2\qq^2}{12\pi^2}\sqrt{1-4m_c^2/s}\, ,
\\
\rho^{\qq\qGqb}_8(s)&=&\frac{m_c^2\qq\qGqa}{24\pi^2}\int_0^1d\alpha
\Bigg\{\frac{m_c^2}{\alpha^2}\delta'\left(s-\tilde{m}^2_c\right)-
\frac{1}{\alpha}\delta\left(s-\tilde{m}^2_c\right)\Bigg\}\, .
\label{SD0++8}
\end{eqnarray}
}

\item For the current $J_9$ with $J^{PC}=0^{++}$:
{\allowdisplaybreaks
\begin{eqnarray}
\nonumber
\rho^{pert}_9(s)&=&\frac{1}{128\pi^6}\dab\frac{(1-\alpha-\beta)^2\left[m_c^2(\alpha+\beta)-3\alpha\beta
s\right]\f(s)^3}{\alpha^3\beta^3}\, , \non
\rho^{\qq}_9(s)&=&\frac{m_c\qq}{4\pi^4}\dab\frac{(1-\alpha-\beta)\f(s)\left[m_c^2(\alpha+\beta)-2\alpha\beta
s\right]}{\alpha^2\beta}\, , \non
\rho^{\GGa}_9(s)&=&\frac{\GGb}{128\pi^6}\dab\Bigg\{\frac{m_c^2(1-\alpha-\beta)^2\left[2m_c^2(\alpha+\beta)-3\alpha\beta
s\right]}{3\beta^3} \non&&
+\frac{\left[m_c^2(\alpha+\beta)-2\alpha\beta
s\right]\f(s)}{8}\left[\frac{(1-\alpha-\beta)^2}{2\alpha^2\beta^2}+\frac{(2-2\alpha-\beta)}{\alpha\beta^2}\right]\Bigg\}\,
, \non \rho^{\qGqb}_9(s)&=&\frac{m_c\qGqa}{64\pi^4}\dab
\frac{(1-5\beta)\left[2m_c^2(\alpha+\beta)-3\alpha\beta
s\right]}{\alpha\beta}\, , \non
\rho^{\qq^2}_9(s)&=&\frac{m_c^2\qq^2}{3\pi^2}\sqrt{1-4m_c^2/s}\, ,
\\
\rho^{\qq\qGqb}_9(s)&=&\frac{m_c^2\qq\qGqa}{24\pi^2}\int_0^1d\alpha
\Bigg\{\frac{4m_c^2}{\alpha^2}\delta'\left(s-\tilde{m}^2_c\right)-
\frac{1}{\alpha}\delta\left(s-\tilde{m}^2_c\right)\Bigg\}\, .
\label{SD0++9}
\end{eqnarray}
}

\item For the current $J_{10}$ with $J^{PC}=0^{++}$:
{\allowdisplaybreaks
\begin{eqnarray}
\nonumber
\rho^{pert}_{10}(s)&=&\frac{3}{64\pi^6}\dab\frac{(1-\alpha-\beta)^2\left[m_c^2(\alpha+\beta)-3\alpha\beta
s\right]\f(s)^3}{\alpha^3\beta^3}\, , \non
\rho^{\qq}_{10}(s)&=&\rho^{\qGqb}_{10}(s)=0\, , \non
\rho^{\GGa}_{10}(s)&=&\frac{\GGb}{64\pi^6}\dab\Bigg\{\frac{m_c^2(1-\alpha-\beta)^2\left[2m_c^2(\alpha+\beta)-3\alpha\beta
s\right]}{\beta^3} \non&&
+\frac{\left[m_c^2(\alpha+\beta)-2\alpha\beta
s\right]\f(s)}{2}\left[\frac{(1-\alpha-\beta)^2}{2\alpha^2\beta^2}+\frac{1}{\alpha\beta}\right]\Bigg\}\,
, \non
\rho^{\qq^2}_{10}(s)&=&\frac{2m_c^2\qq^2}{\pi^2}\sqrt{1-4m_c^2/s}\,
,
\\
\rho^{\qq\qGqb}_{10}(s)&=&\frac{m_c^2\qq\qGqa}{\pi^2}\int_0^1d\alpha
\frac{m_c^2}{\alpha^2}\delta'\left(s-\tilde{m}^2_c\right)\, .
\label{SD0++10}
\end{eqnarray}
}

\item For the current $J_{11\mu\nu}$ with $J^{PC}=2^{++}$:
{\allowdisplaybreaks
\begin{eqnarray}
\nonumber \rho^{pert}_{11}(s)&=&\frac{1}{384\pi^6}\dab \non
&&\Bigg\{\frac{(1-\alpha-\beta)^3\left[m_c^2(\alpha+\beta)-5\alpha\beta
s\right]\left[3m_c^2(\alpha+\beta)-17\alpha\beta
s\right]\f(s)^2}{2\alpha^3\beta^3} \non
&&-(1-\alpha-\beta)^2\f(s)^3\left[\frac{13(1-\alpha-\beta)s}{\alpha^2\beta^2}-\frac{17m_c^2(\alpha+\beta)-81\alpha\beta
s}{2\alpha^3\beta^3}\right]\Bigg\}\, , \non
\rho^{\qq}_{11}(s)&=&-\frac{5m_c\qq}{2\pi^4}\dab\frac{(1-\alpha-\beta)\f(s)\left[m_c^2(\alpha+\beta)-3\alpha\beta
s\right]}{\alpha^2\beta}\, , \non
\rho^{\GGa}_{11}(s)&=&\frac{\GGb}{384\pi^6}\dab\Bigg\{\frac{m_c^2(1-\alpha-\beta)^3\left[3m_c^2(\alpha+\beta)-16\alpha\beta
s\right]}{3\alpha^3} \non&&
+\frac{m_c^2(1-\alpha-\beta)^2\left[17m_c^2(\alpha+\beta)-33\alpha\beta
s\right]}{3\alpha^3}-\frac{5\left[m_c^2(\alpha+\beta)-3\alpha\beta
s\right]\f(s)}{4\alpha\beta} \non &&
-\frac{5(1-\alpha-\beta)^2\left[5m_c^2(\alpha+\beta)-13\alpha\beta
s\right]\f(s)}{48\alpha^2\beta^2} \non &&
-\frac{(1-\alpha-\beta)\left[29m_c^2(\alpha+\beta)-88\alpha\beta
s\right]\f(s)}{3\alpha^2\beta} \non &&
-\frac{5(1-\alpha-\beta)^3}{144\alpha^2\beta^2}\left[4(\alpha\beta
s)^2-18\alpha\beta s\f(s)+3\f(s)^2\right] \non &&
+\frac{(1-\alpha-\beta)^2}{12\alpha^2\beta}\left[28(\alpha\beta
s)^2-78\alpha\beta s\f(s)+9\f(s)^2\right]\Bigg\} \non &&
+\frac{7m_c^6\GGb}{1728\pi^6}\int^{\alpha_{max}}_{\alpha_{min}}d\alpha\frac{s^2(1-\alpha)\left[m_c^2-\alpha(1-\alpha)s\right]^3}{\left[m_c^2-(1-\alpha)s\right]^6}\,
, \non \rho^{\qGqb}_{11}(s)&=&\frac{5m_c\qGqa}{48\pi^4}\dab
\frac{(1+12\alpha-\beta)\left[m_c^2(\alpha+\beta)-2\alpha\beta
s\right]}{\alpha\beta}\, , \non
\rho^{\qq^2}_{11}(s)&=&\frac{10m_c^2\qq^2}{3\pi^2}\sqrt{1-4m_c^2/s}\,
,
\\
\rho^{\qq\qGqb}_{11}(s)&=&\frac{5m_c^2\qq\qGqa}{36\pi^2}\int_0^1d\alpha
\Bigg\{\frac{12m_c^2}{\alpha^2}\delta'\left(s-\tilde{m}^2_c\right)+
\frac{1}{\alpha}\delta\left(s-\tilde{m}^2_c\right)\Bigg\}\, .
\label{SD2++11}
\end{eqnarray}
}

\item For the current $J_{12\mu\nu}$ with $J^{PC}=2^{++}$:
{\allowdisplaybreaks
\begin{eqnarray}
\nonumber \rho^{pert}_{12}(s)&=&\frac{1}{384\pi^6}\dab \non
&&\Bigg\{\frac{(1-\alpha-\beta)^3\left[m_c^2(\alpha+\beta)-5\alpha\beta
s\right]\left[3m_c^2(\alpha+\beta)-17\alpha\beta
s\right]\f(s)^2}{2\alpha^3\beta^3} \non
&&-(1-\alpha-\beta)^2\f(s)^3\left[\frac{13(1-\alpha-\beta)s}{\alpha^2\beta^2}-\frac{17m_c^2(\alpha+\beta)-81\alpha\beta
s}{2\alpha^3\beta^3}\right]\Bigg\}\, , \non
\rho^{\qq}_{12}(s)&=&\frac{5m_c\qq}{2\pi^4}\dab\frac{(1-\alpha-\beta)\f(s)\left[m_c^2(\alpha+\beta)-3\alpha\beta
s\right]}{\alpha^2\beta}\, , \non
\rho^{\GGa}_{12}(s)&=&\frac{\GGb}{384\pi^6}\dab\Bigg\{\frac{m_c^2(1-\alpha-\beta)^3\left[3m_c^2(\alpha+\beta)-16\alpha\beta
s\right]}{3\alpha^3} \non&&
+\frac{m_c^2(1-\alpha-\beta)^2\left[17m_c^2(\alpha+\beta)-33\alpha\beta
s\right]}{3\alpha^3}-\frac{5\left[m_c^2(\alpha+\beta)-3\alpha\beta
s\right]\f(s)}{4\alpha\beta} \non &&
-\frac{5(1-\alpha-\beta)^2\left[5m_c^2(\alpha+\beta)-13\alpha\beta
s\right]\f(s)}{48\alpha^2\beta^2} \non &&
-\frac{(1-\alpha-\beta)\left[29m_c^2(\alpha+\beta)-88\alpha\beta
s\right]\f(s)}{3\alpha^2\beta} \non &&
-\frac{5(1-\alpha-\beta)^3}{144\alpha^2\beta^2}\left[4(\alpha\beta
s)^2-18\alpha\beta s\f(s)+3\f(s)^2\right] \non &&
+\frac{(1-\alpha-\beta)^2}{12\alpha^2\beta}\left[28(\alpha\beta
s)^2-78\alpha\beta s\f(s)+9\f(s)^2\right]\Bigg\} \non &&
+\frac{7m_c^6\GGb}{1728\pi^6}\int^{\alpha_{max}}_{\alpha_{min}}d\alpha\frac{s^2(1-\alpha)\left[m_c^2-\alpha(1-\alpha)s\right]^3}{\left[m_c^2-(1-\alpha)s\right]^6}\,
, \non \rho^{\qGqb}_{12}(s)&=&-\frac{5m_c\qGqa}{48\pi^4}\dab
\frac{(1+12\alpha-\beta)\left[m_c^2(\alpha+\beta)-2\alpha\beta
s\right]}{\alpha\beta}\, , \non
\rho^{\qq^2}_{12}(s)&=&\frac{10m_c^2\qq^2}{3\pi^2}\sqrt{1-4m_c^2/s}\,
,
\\
\rho^{\qq\qGqb}_{12}(s)&=&\frac{5m_c^2\qq\qGqa}{36\pi^2}\int_0^1d\alpha
\Bigg\{\frac{12m_c^2}{\alpha^2}\delta'\left(s-\tilde{m}^2_c\right)+
\frac{1}{\alpha}\delta\left(s-\tilde{m}^2_c\right)\Bigg\}\, .
\label{SD2++12}
\end{eqnarray}
}

\item For the current $J_{13\mu\nu}$ with $J^{PC}=2^{++}$:
{\allowdisplaybreaks
\begin{eqnarray}
\nonumber \rho^{pert}_{13}(s)&=&\frac{1}{192\pi^6}\dab \non
&&\Bigg\{\frac{(1-\alpha-\beta)^3\left[m_c^2(\alpha+\beta)-5\alpha\beta
s\right]\left[3m_c^2(\alpha+\beta)-17\alpha\beta
s\right]\f(s)^2}{2\alpha^3\beta^3} \non
&&-(1-\alpha-\beta)^2\f(s)^3\left[\frac{13(1-\alpha-\beta)s}{\alpha^2\beta^2}-\frac{17m_c^2(\alpha+\beta)-81\alpha\beta
s}{2\alpha^3\beta^3}\right]\Bigg\}\, , \non
\rho^{\qq}_{13}(s)&=&-\frac{5m_c\qq}{\pi^4}\dab\frac{(1-\alpha-\beta)\f(s)\left[m_c^2(\alpha+\beta)-3\alpha\beta
s\right]}{\alpha^2\beta}\, , \non
\rho^{\GGa}_{13}(s)&=&\frac{\GGb}{576\pi^6}\dab\Bigg\{\frac{m_c^2(1-\alpha-\beta)^3\left[3m_c^2(\alpha+\beta)-16\alpha\beta
s\right]}{\alpha^3} \non&&
+\frac{m_c^2(1-\alpha-\beta)^2\left[17m_c^2(\alpha+\beta)-33\alpha\beta
s\right]}{\alpha^3}-\frac{25\left[m_c^2(\alpha+\beta)-2\alpha\beta
s\right]\f(s)}{4\alpha\beta} \non &&
-\frac{25(1-\alpha-\beta)^2\left[5m_c^2(\alpha+\beta)-13\alpha\beta
s\right]\f(s)}{32\alpha^2\beta^2} \non &&
-\frac{(1-\alpha-\beta)\left[16m_c^2(\alpha+\beta)-47\alpha\beta
s\right]\f(s)}{2\alpha^2\beta} \non &&
-\frac{25(1-\alpha-\beta)^3}{96\alpha^2\beta^2}\left[4(\alpha\beta
s)^2-18\alpha\beta s\f(s)+3\f(s)^2\right] \non &&
-\frac{(1-\alpha-\beta)^2}{8\alpha^2\beta}\left[28(\alpha\beta
s)^2-78\alpha\beta s\f(s)+9\f(s)^2\right]\Bigg\} \non &&
+\frac{7m_c^6\GGb}{864\pi^6}\int^{\alpha_{max}}_{\alpha_{min}}d\alpha\frac{s^2(1-\alpha)\left[m_c^2-\alpha(1-\alpha)s\right]^3}{\left[m_c^2-(1-\alpha)s\right]^6}\,
, \non \rho^{\qGqb}_{13}(s)&=&\frac{5m_c\qGqa}{48\pi^4}\dab
\frac{(5+24\alpha-5\beta)\left[m_c^2(\alpha+\beta)-2\alpha\beta
s\right]}{\alpha\beta}\, , \non
\rho^{\qq^2}_{13}(s)&=&\frac{20m_c^2\qq^2}{3\pi^2}\sqrt{1-4m_c^2/s}\,
,
\\
\rho^{\qq\qGqb}_{13}(s)&=&\frac{5m_c^2\qq\qGqa}{36\pi^2}\int_0^1d\alpha
\Bigg\{\frac{24m_c^2}{\alpha^2}\delta'\left(s-\tilde{m}^2_c\right)+
\frac{5}{\alpha}\delta\left(s-\tilde{m}^2_c\right)\Bigg\}\, .
\label{SD2++13}
\end{eqnarray}
}

\item For the current $J_{14\mu\nu}$ with $J^{PC}=2^{++}$:
{\allowdisplaybreaks
\begin{eqnarray}
\nonumber \rho^{pert}_{14}(s)&=&\frac{1}{192\pi^6}\dab \non
&&\Bigg\{\frac{(1-\alpha-\beta)^3\left[m_c^2(\alpha+\beta)-5\alpha\beta
s\right]\left[3m_c^2(\alpha+\beta)-17\alpha\beta
s\right]\f(s)^2}{2\alpha^3\beta^3} \non
&&-(1-\alpha-\beta)^2\f(s)^3\left[\frac{13(1-\alpha-\beta)s}{\alpha^2\beta^2}-\frac{17m_c^2(\alpha+\beta)-81\alpha\beta
s}{2\alpha^3\beta^3}\right]\Bigg\}\, , \non
\rho^{\qq}_{14}(s)&=&\frac{5m_c\qq}{\pi^4}\dab\frac{(1-\alpha-\beta)\f(s)\left[m_c^2(\alpha+\beta)-3\alpha\beta
s\right]}{\alpha^2\beta}\, , \non
\rho^{\GGa}_{14}(s)&=&\frac{\GGb}{576\pi^6}\dab\Bigg\{\frac{m_c^2(1-\alpha-\beta)^3\left[3m_c^2(\alpha+\beta)-16\alpha\beta
s\right]}{\alpha^3} \non&&
+\frac{m_c^2(1-\alpha-\beta)^2\left[17m_c^2(\alpha+\beta)-33\alpha\beta
s\right]}{\alpha^3}-\frac{25\left[m_c^2(\alpha+\beta)-2\alpha\beta
s\right]\f(s)}{4\alpha\beta} \non &&
-\frac{25(1-\alpha-\beta)^2\left[5m_c^2(\alpha+\beta)-13\alpha\beta
s\right]\f(s)}{32\alpha^2\beta^2} \non &&
-\frac{(1-\alpha-\beta)\left[16m_c^2(\alpha+\beta)-47\alpha\beta
s\right]\f(s)}{2\alpha^2\beta} \non &&
-\frac{25(1-\alpha-\beta)^3}{96\alpha^2\beta^2}\left[4(\alpha\beta
s)^2-18\alpha\beta s\f(s)+3\f(s)^2\right] \non &&
-\frac{(1-\alpha-\beta)^2}{8\alpha^2\beta}\left[28(\alpha\beta
s)^2-78\alpha\beta s\f(s)+9\f(s)^2\right]\Bigg\} \non &&
+\frac{7m_c^6\GGb}{864\pi^6}\int^{\alpha_{max}}_{\alpha_{min}}d\alpha\frac{s^2(1-\alpha)\left[m_c^2-\alpha(1-\alpha)s\right]^3}{\left[m_c^2-(1-\alpha)s\right]^6}\,
, \non \rho^{\qGqb}_{14}(s)&=&-\frac{5m_c\qGqa}{48\pi^4}\dab
\frac{(5+24\alpha-5\beta)\left[m_c^2(\alpha+\beta)-2\alpha\beta
s\right]}{\alpha\beta}\, , \non
\rho^{\qq^2}_{14}(s)&=&\frac{20m_c^2\qq^2}{3\pi^2}\sqrt{1-4m_c^2/s}\,
,
\\
\rho^{\qq\qGqb}_{14}(s)&=&\frac{5m_c^2\qq\qGqa}{36\pi^2}\int_0^1d\alpha
\Bigg\{\frac{24m_c^2}{\alpha^2}\delta'\left(s-\tilde{m}^2_c\right)+
\frac{5}{\alpha}\delta\left(s-\tilde{m}^2_c\right)\Bigg\}\, .
\label{SD2++14}
\end{eqnarray}
}

\end{itemize}

\end{document}